\documentclass[eqsecnum,showpacs,preprint,aps]{revtex4}
\usepackage{graphics}
\def\beq{\begin{equation}}
\def\eeq{\end{equation}}
\def\bea{\begin{eqnarray}}
\def\eea{\end{eqnarray}}
\def\bwi{\begin{widetext}}
\def\ewi{\end{widetext}}
\def\lf{\left}
\def\rt{\right}

\def\tst{\textstyle}

\def\fno#1{Fig.~\ref{#1}}

\def\eno#1{Eq.~(\ref{#1})}
\def\Eno#1{Equation (\ref{#1})}
\def\etwo#1#2{Eqs.~(\ref{#1}) and (\ref{#2})}

\def\al{\alpha}

\def\gam{\gamma}

\def\tta{\theta}

\def\lam{\lambda}

\def\om{\omega}

\def\sech{{\rm sech\,}}
\def\apx{\approx}
\def\ptl{\partial}
\def\by{\over}

\def\hf{{1\over2}}
\def\tshf{\tst\hf}

\def\ham{{\cal H}}
\def\ket#1{|#1\rangle}

\def\tran#1#2{\langle#1|#2\rangle}

\def\mel#1#2#3{\langle#1|#2|#3\rangle}

\def\bH{{\bf H}}
\def\bJ{{\bf J}}

\def\zhat{{\bf{\hat z}}}
\def\nhat{{\bf{\hat n}}}

\def\Fe8{Fe$_8$}

\def\rtl{\sqrt{\lam}}

\def\zbar{{\bar z}}
\def\baz{{\bar z}}

\def\rtl{\sqrt\lam}

\def\rtb{\sqrt{1-\htil^2}}

\def\htil{h}

\begin{document}


\title{Large-field versus discontinuous instantons in spin orientation tunneling}

\author{Anupam Garg}
\email[e-mail address: ]{agarg@northwestern.edu}
\author{Ersin Ke\c{c}ecio\u{g}lu}
\affiliation{Department of Physics and Astronomy, Northwestern University,
Evanston, Illinois 60208}

\date{\today}

\begin{abstract}
Tunnel splitting oscillations in magnetic molecules are reconsidered
within the simplest model for the problem, which does not contain fourth
order anisotropy. It is shown that at large magnetic field, there is only
one instanton, and it is continuous. This is in contrast to the
discontinuous instantons that are induced by the fourth order term
[Ersin Ke\c{c}ecio\u{g}lu and A.~Garg, Phys. Rev.
Lett. {\bf 88}, 237205 (2002)].
\end{abstract}


\maketitle

\section{Introduction}
The purpose of this short note is to clarify the differences
between certain types of instantons that arise in the study of
spin orientation tunneling in molecular magnets.
The problem derives from the specific phenomenon of tunnel splitting or
gap oscillations in
the molecular ion [(tacn)$_6$Fe$_8$O$_2$(OH)$_{12}$]$^{8+}$  
(abbreviated to \Fe8 henceforth) \cite{werns}, but our considerations
should apply to other molecular magnetic system as well.

The simplest model Hamiltonian that leads to gap oscillations is
\beq
\ham_0 = k_1 J_z^2 + k_2 J_y^2 - g\mu_B \bJ\cdot\bH, \label{ham0}
\eeq
where $\bJ$ is a dimensionless spin operator, $k_1 > k_2 > 0$, and
$\bH$ is an external magnetic field. It is known that for
$\bH \| \zhat$, the tunnel splitting, or gap, between the two lowest eigen
states of $\ham_0$ oscillates as a function of $H_z$, vanishing 
whenever \cite{gargepl}
\beq
H_z = {1 \by J}\left(J - n - \tshf \right) H_c.
   \label{quench0}
\eeq
Here,
\beq
\lam = k_2/k_1, {\rm\ and\ } H_c = 2k_1J/g\mu_B.
\eeq
For \Fe8, for which $J = 10$, $k_1 = 0.338$~K, and $k_2 = 0.246$~K,
this would imply that the splitting is quenched at 10 values of $H_z > 0$,
equally spaced by 0.263 T. This quenching phenomenon is nicely
interpreted in terms of interfering semiclassical paths, or
instantons \cite{wilk}.

In reality, only four quenches are seen in \Fe8, and the spacing between
the quenches is $\sim 0.41$~T, 50\% larger than predicted by \eno{quench0}.
Both these facts are accounted for by a slightly modified Hamiltonian
\beq
\ham_1 = \ham_0 - C [(J_z + iJ_y)^4 + {\rm h.c.}], \label{ham1}
\eeq
with $C = 0.29\,\mu$K, as is easily verified by explicit numerical
diagonalization. The fact that only 4 quenches are seen is paradoxical,
however. The relative phase between the interfering instantons is a
topological quantity, so that it must equal $J\pi$ at $H=0$, and tend to
0 as $H \to H_c$. There is no apparent way to get other than 10
quenches. The resolution is that the $C$ term, though small, has
a dramatic effect: any non zero $C$ leads to the existence of instantons
that are {\it discontinuous\/} at the end points \cite{kg2}. One of
these instantons ends up being the dominant one by virtue of having the
lowest action beyond a certain field ($\sim 0.25H_c$ for \Fe8), and since it
has no interfering partner, there are no more quenches of the splitting
beyond this field. In this way, the discontinuous or jump instantons may
be said to shunt the interference effect. The jump instantons have
discontinuities, and thus do not have to respect the topology.

In private correspondence to us, E.~Chudnovsky has alleged that
jump instantons were first introduced by him and Hidalgo in an earlier
study of the simpler Hamiltonian (\ref{ham0}) at high fields \cite{ecxmh}.
We disagree. The instanton studied by Chudnovsky and Hidalgo (CH) is of
exactly the standard continuous type arising in the study of a particle tunneling in
a symmetric one-dimensional double well, but its nature is obscured by
the subtleties of spin trajectories, and the singularities of the spherical
polar coordinate system on the sphere. These subtleties are vexing and the
source of innumerable pitfalls, so it seems useful to us to elaborate
on the instantons in question, and show explicitly that they are continuous.
We also use the opportunity to correct a misconception in a still
earlier study of the same problem, i.e., \eno{ham0} at high fields,
by one of us \cite{agprb99}.

\section{Details}
We consider only fields parallel to $\zhat$, and
define the reduced field
\beq
h = H/H_c,
\eeq
and the special value
\beq
h^* = (1-\lam)^{1/2}.
\eeq
CH study the problem for $h > h^*$, and Chudnovsky is asserting that
the instantons in that study have jumps. We give three arguments to
show that this is not so.

The first and mathematically simplest way is to consider fields
just less than $H_c$. We write
\beq
h = (1 - \eta)
\eeq
where $0 < \eta \ll 1$. The expectation value of $J_z$ in the low energy
states is then extremely close to $J$, so we can approximate
\beq
J_x \apx J q, \quad J_y \apx p,
\eeq
where $p$ and $q$ are momentum and position operators with the
commutator $[q,p] = i$. It is self-consistently verifiable that
$J_y \sim \eta$, $J_x \sim \eta^{1/2}$, so that we can write
$J_z \simeq (J^2 - J_x^2 - J_y^2)^{1/2}$. Expanding the term $-g\mu_bH J_z$
to order $J_y^2$ and $J_x^4$, rewriting in terms of $q$ and $p$, and omitting an
additive constant, we arrive at the equivalent Hamiltonian
\beq
\ham_{\rm equiv} = {1\by 2m} p^2 + {1\by 4} k_1 h J^2 q^4
                                        - \eta k_1 J^2 q^2.
                    \label{hamquar}
\eeq
Here,
\beq
m = [2(k_2 - \eta k_1)]^{-1} \apx (2k_2)^{-1}.
\eeq

\Eno{hamquar} is the Hamiltonian for the particle in a quartic double
well. The minima of the well are at $\pm q_0 = \pm (2\eta)^{1/2}.$
The instantons for this problem are known, and the one that runs from
$q_0$ to $-q_0$ is given by
\bea
q(t) &=& -q_0 \tanh\gam t, \label{quarq}\\
p(t) &=& -i m\gam q_0 \tanh\gam t \,\sech^2 \gam t. \label{quarp}
\eea
Here, $t$ is an imaginary time, and $\gam = (2k_1 k_2 \eta)^{1/2} J$.
If we transcribe these results back into $J_x$, $J_y$, and $J_z$, it is
manifest that the spin trajectory goes {\it continuously\/} from one
minimum of the classical energy to the other without any jumps. There is
nothing in the physical character of the problem to indicate a qualitative
change as $H$ is reduced, so as long as one is in the field range where
the gap does not oscillate ($h> h^*$), the instantons should be continuous.

In our second argument,
we study the instantons for all $H$. Since these lie on the complex
unit sphere, it is highly advantageous to use stereographic coordinates,
\beq
z = \tan\tshf\tta e^{i\phi}, \quad
\zbar = \tan\tshf\tta e^{-i\phi}, 
\eeq
where $\tta$ and $\phi$ are the customary spherical polar coordinates.
If a direction $\nhat$ has coordinates $(\tta, \phi)$, and if 
$\ket\nhat$ is the spin coherent state with maximal spin projection along
$\nhat$, then up to normalization and phase, $\ket\nhat$ is identical
to $\ket z \equiv e^{zJ_-}\ket{J,J}$ where $\bJ^2\ket{J,J} = J(J+1)\ket{J,J}$
and $J_z \ket{J,J} = J \ket{J,J}$. A point on the complex unit sphere
is parametrized by complex $\tta$ and $\phi$, which means that $z$ and
$\zbar$ {\it need not\/} be complex conjugates; rather, they are
independent complex variables. Either coordinate system shows that the
complex unit sphere is a four-dimensional manifold, but the $(z,\zbar)$
system handles coordinate singularities better. In terms of
$z$ and $\zbar$, the cartesian components of $\bJ$ are given by
\beq
(J_x, J_y, J_z) = {J \by 1 + \zbar z}
                      \bigl(z + \zbar, -i(z - \zbar), 1 - \zbar z \bigr).
   \label{Jxyz}
\eeq
The south pole is the point $z = \zbar = \infty$; if $z \to \infty$ with
$\zbar$ finite, $\bJ \to J(\zbar^{-1}, -i\zbar^{-1}, -1)$; if
$\zbar \to \infty$ with $z$ finite, $\bJ \to J(z^{-1}, i z^{-1}, -1)$.

We have found the instantons for this problem in Ref.~\onlinecite{gkps}
among other places, but we restate the results along with some attendant
formulas for ready reference, and to enable readers to verify the
key results for themselves. Let us define
\beq
E({\bar z}, z) = \mel{z}{\ham}{z}/ \tran{z}{z}.
\eeq
Then the Euler-Lagrange equations of motion for the instantons
are given by
\beq
\dot\baz = {(1 +\baz z)^2 \over 2J} {\ptl E \over \ptl z}, \qquad
\dot z = -{(1 +\baz z)^2 \over 2J} {\ptl E \over \ptl\baz}. \label{ELzz}
\eeq

With the Hamiltonian (\ref{ham0}),
\beq
E(\baz,z) = k_1 J^2
            \lf[(1-\baz z)^2 - \lam (z - \baz)^2 - 2h (1 - \baz^2 z^2)
                   \over
                 (1 + \baz z)^2 \rt], \label{hamfe8}
\eeq
so that 
\bwi
\bea
{\dot z} &=& -{k_1 J \by (1 + \zbar z)}
            \left[-2 z (1 - \zbar z) + \lam (z - \zbar)(1+z^2)
                           + 2 h z (1 + \zbar z) \right], \label{eomz}\\
{\dot \zbar} &=& {k_1 J \by (1 + \zbar z)}
            \left[-2\zbar (1 - \zbar z) + \lam (\zbar - z)(1+\zbar^2)
                           + 2 h \zbar (1 + \zbar z) \right]. \label{eomzb}
\eea
\ewi
The minima of $E$ are at $\baz = z = \pm z_0$ where
$z_0 = [(1-h)/(1+h)]^{1/2}$. There are two solutions to the equations of
motion that run from $z_0$ to $-z_0$. The first is given by
\bea
z^{(1)}(t) &=& -z_0 \tanh \tau,  \label{EQ:zcl} \label{zt} \\
\zbar^{(1)}(t) &=& -z_0 { \rtl\tanh \tau + \rtb 
                       \over \rtl + \rtb\tanh \tau}, \label{zbart}
\eea
while the second is given by
\bea
z^{(2)}(t) &=& -z_0 { \rtl\tanh \tau - \rtb 
                       \over \rtl - \rtb\tanh \tau},  \label{EQ:zcl2} \\
\zbar^{(2)}(t) &=& -z_0 \tanh \tau.                   \label{EQ:bzcl2}
\eea
We have defined $\tau = \om t/2$, with
$\om = 2k_1 J[\lam(1-h^2)]^{1/2}$. Readers can verify by direct substitution
into the equations of motion that these are indeed solutions. Also, we have
\beq
z^{(2)}(t) = - \zbar^{(1)}(-t), \quad
\zbar^{(2)}(t) = - z^{(1)}(-t).
\eeq

For $h < h^*$, the two solutions can be seen to be distinct by noting that
in the first $\zbar$ diverges at some $t$ with $z$ remaining finite,
while in the other the converse happens. The action for these two instantons
differs by a field dependent phase, which gives rise to the gap oscillation.

For $h > h^*$, on the other hand, the two solutions are physically identical.
To see this, let us note that we can write
\bea
z^{(1)} = - z_0 \tanh \tau,&& \ 
          \zbar^{(1)} = -z_0 \tanh(\tau + \tau_0), \\
z^{(2)} = -z_0 \tanh(\tau - \tau_0),&& \ 
\zbar^{(2)} = - z_0 \tanh \tau,
\eea
where,
\beq
\tau_0 = \tanh^{-1}\sqrt{1-h^2 \by \lam}.
\eeq
If $h > h^*$, $\tau_0$ is real, so the two solutions are related by a simple
translation in time. Such translations are automatically included in the
sum over multiinstanton paths, so there is only one type of instanton.

More importantly, the instanton is entirely continuous.
We plot the components of $\bJ$ versus $\tau$ in \fno{jcomps}. As can be seen,
there is no discontinuity in any component. This figure should be compared
with Fig.~2 of Ref.~\onlinecite{kg2}, which pertains to the problem with
the $C$ term in \eno{ham1}, and shows instantons with patent jumps.

\begin{figure}
\includegraphics{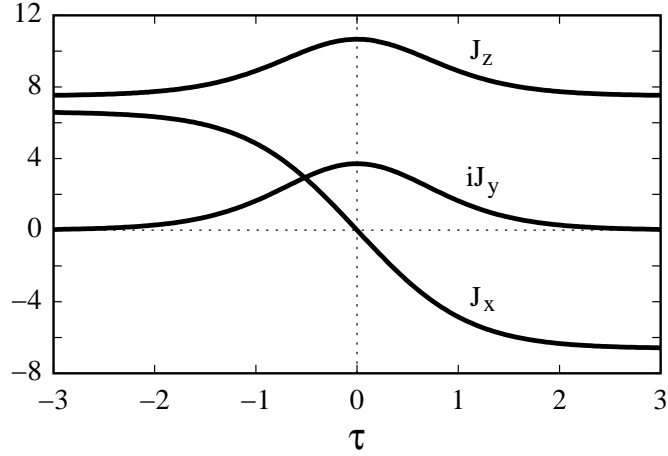}
\caption{\label{jcomps} Components of the complex vector $\bJ(\tau)$
for the instanton solution, versus $\tau$.
The {\it x\/} and {\it z\/} components are real, the {\it y\/} component is
pure imaginary.}
\end{figure}

Thirdly, let us consider the problem in spherical polar coordinates, as
is also done by CH, and in Ref.~\onlinecite{agprb99}. CH proceed by
integrating out $\cos\tta$, which is feasible as the integral is Gaussian.
The resulting action for $\phi$ is like that of a massive particle. If one
now seeks instantons for this problem analogously to the particle in a
quartic double well say, {\it and insists on a solution $\phi(\tau)$ in
which $\phi$ is real\/}, one discovers that the variable $\tau$ cannot be real,
but rather must follow a Z-shaped contour in the complex plane, consisting of
two segments parallel to the real axis, joined by a segment along the
imaginary axis \cite{complext}. This is very different, however, from a
discontinuity in $\phi$
itself. Indeed, CH seem aware of this point, for they say {\it ``It is clear
from the shape of the potential that all three parts of the trajectory
join smoothly, because $\phi(\tau)$ and $\dot\phi(\tau)$ coincide at the
joints."\/} We agree with this. And, it is clear that if one finds the
dominant path for $\cos\tta(\tau)$, i.e., the value of $\cos\tta$ for which
the original Gaussian factor is maximal, one will discover that $\cos\tta(\tau)$
is also continuous. To eliminate any doubt, we show this by examining
the explicit solution for the instanton from Ref. \onlinecite{agprb99}.
In our present notation, this reads
\bwi
\bea
\cos\phi &=& -i {(h^2 + \lam -1)^{1/2} \tanh2\tau
                   \over (1 - h^2 - \lam \tanh^2 2\tau)^{1/2}} \\
\cos\tta &=& {h(1-h^2 - \lam\tanh^2 2\tau)
                 - (1-h^2) \lam^{1/2} (h^2 + \lam - 1)^{1/2} \sech2\tau
                      \over
               1 - h^2 - \lam + h^2\lam\,\sech^2 2\tau}.
\eea
\ewi
From these one can construct the combinations $\tan\hf\tta e^{\pm i\phi}$
to obtain $z$ and $\zbar$. After a certain amount of algebra, we obtain
\beq
\left( \begin{array}{c} z(t) \\ \zbar(t) \end{array} \right)
  = -z_0 {\tanh\tau \pm \al \over 1 \pm \al \tanh\tau}, \label{inst2}
\eeq
where
\beq
\al = {(h^2 + \lam - 1)^{1/2} - \lam^{1/2} \over (1-h^2)^{1/2}}.
\eeq
If we now note that
\beq
\tanh^{-1}\al = - \tau_0/2,
\eeq
we see that \eno{inst2} is the same as \etwo{zt}{zbart} up to a shift
in $\tau$. Again, we see that there is no discontinuity in the instanton.

It is a separate issue as to what contour what chooses for $\tau$.
If, instead of
the real line, one lets $\tau$ run on the Z-shaped contour in 
Ref.~\onlinecite{agprb99}, $z$ will run from $z_0$ to $-z_0$ along some
contour other than the real line, while $\zbar$ will run along some other
contour, such that
\beq
\zbar(t) = {\rtl z(t) - (1-h) \over \rtl - (1+h) z(t)} \label{zbar(z)}
\eeq
[This can be verified directly from \eno{inst2}, and it is a consequence
of energy conservation along the instanton.] Since the kinetic term in
the action can be written as a contour integral over $z$,
\beq
S = J \int {1 \by 1 + \zbar z}
             \left[ \zbar(z) - z {d \zbar \by dz} \right] dz
\eeq
it follows that any deformation of the $z$-contour that does not cross any
singularities of $\zbar(z)$ is allowed. This is indeed the case for us, as
for $h > h^*$, the pole at $z = \rtl/(1+h)$ lies outside the segment
$[-z_0,z_0]$ on the real line.

We conclude by clarifying the misconception in Ref.~\onlinecite{agprb99}.
The discussion
above shows that there is only one instanton when $h > h^*$, whereas, in the
earlier work it is implicit that there are two, with equal, real Euclidean
actions \cite{chact2}. However, the central thesis of that work, and the
calculation of the action itself, are completely unaffected.

One further point is that at exactly at $h = h^*$, standard instanton
calculations of the Hamiltonian (\ref{ham0}) cease to be valid, because two
distinct instantons merge into one. This is rather like the phenomenon of
merging critical points in the steepest descent approximation for ordinary
integrals \cite{bleihan}. We expect existing formulas for the tunnel splitting 
to be inaccurate in a narrow window of fields near $h=h^*$. The gap
can undoubtedly be bridged via a uniform asymptotic approximation, which
could be developed using the discrete phase integral method as in
Ref.~\onlinecite{agdpi}. In the methodology of that paper, what happens
as we cross the point $h = h^*$ from below is that the width of the unusual
forbidden region, in which the wavefunctions are exponentially damped with
oscillations, shrinks to zero. In this case, one should join the wavefunctions
on the two sides of this region using one quadratic turning point connection
formula as opposed to two linear turning point formulas. We leave this as an
exercise for the reader.

In summary, we have shown that that the high-field instantons in the 
\Fe8 problem are continuous, and thus qualititatively distinct from the
discontinuous instantons induced by the fourth order anisotropy in
\eno{ham1}.

\section{Acknowledgments}
This research is supported by the National Science Foundation via Grant
No. DMR-0202165.

%

%
%

\begin{thebibliography}{99}
\bibitem{chact2} The same misconception appears to be present in
CH, as evidenced by their Eq.~(19), and the surrounding terms, and
their statements that there is an interference phenomenon at fields
greater than $H^*$.
\end{thebibliography}
\end{document}